# Enhanced magnetoresistance in perpendicular magnetic tunneling junctions with MgAl$_2$O$_4$ barrier

Pravin Khanal, Bowei Zhou, Magda Andrade, Christopher Mastrangelo, Ali Habiboglu, Arthur Enriquez, Daulton Fox, Kennedy Warrilow, and Wei-Gang Wang[*]

Department of Physics, University of Arizona, Tucson, AZ 85721, USA

Perpendicular magnetic tunnel junction with MgAl$_2$O$_4$ barrier is investigated. It is found that reactive RF sputtering with O$_2$ is essential to obtain strong perpendicular magnetic anisotropy and large tunneling magnetoresistance in MgAl$_2$O$_4$-based junctions. An interfacial perpendicular magnetic anisotropy energy density of 2.25 mJ/m$^2$ is obtained for the samples annealed at 400˚C. An enhanced magnetoresistance of 60% has also been achieved. The V$_{half}$, bias voltage at which tunneling magnetoresistance drops to half of the zero-bias value, is found to be about 1V, which is substantially higher than that of MgO-based junctions.

[*]wgwang@physics.arizona.edu

Magnetic tunnel junction (MTJ) is one of the most important devices for magnetic random-access memory(MRAM).[1,2] A very high tunneling magnetoresistance (TMR) has been predicted due to the spin dependent coherent tunneling through the MgO barrier in Fe/MgO/Fe structure.[3,4] Indeed, large TMR has been achieved experimentally in MgO-based junction with Fe,[5] CoFe,[6] and CoFeB,[7] electrodes. Particularly, CoFeB/MgO/CoFeB MTJs have attracted a great deal of attention due to the perpendicular magnetic anisotropy (PMA),[8] and the voltage controlled magnetic anisotropy (VCMA),[9] effect observed in this system. Energy efficient switching in the level of a few femto-Joules have been demonstrated.[10,11] However, one shortcoming of the CoFeB-MgO MTJ is the strong bias voltage dependence of TMR, where TMR quickly drops to very small values when the bias is larger than 1V.[12,13] This is particularly important to the VCMA-induced switching compared to the spin-transfer-torque (STT)-based switching, as a sizable TMR is needed in order to accurately determine the VCMA coefficient[14] at the high switching voltages (>2V).[10,11] It has been suggested that the voltage dependence of TMR is related to the interface quality between the barrier and the ferromagnetic (FM) electrode.[15,16] In addition, the lattice parameters of MgO cannot be tuned to match with other ferromagnetic materials. Therefore, the exploration of other barriers with a weaker bias dependence of TMR and tunable lattice parameter is important to design the energy efficient MTJs.

Various spinel-based oxides have been purposed as an alternative to the MgO barrier such as $MgGa_2O_4$,[17] $MgAl_2O_4$,[15] $GaO_x$/MgO,[18] and $GaO_x$.[19] Among them, the $MgAl_2O_4$ (MAO) barrier has a technological advantage since it can be deposited on the top of an amorphous CoFeB layer without any particular underlayer structure. The (001)-oriented MAO also has a very small lattice mismatch with Fe(001) (~0.3%) and exhibits a spin dependent coherent tunneling similar to the MgO barrier.[20,21] A fully spin polarized $\Delta_1$ state in Fe/Co electrodes couples with the $\Delta_1$ evanescent state of MAO barrier and tunnels through it. Due to a very small lattice mismatch, high TMR ratio and better bias dependence of TMR have been achieved in MAO-MTJ with in-plane magnetic anisotropy.[15,20,22] Theoretical study has showed that the band fold on Fe is responsible for the higher $V_{half}$ in Fe/MAO/Fe structure.[23] Very interesting resonant tunneling through multi-metallic quantum well states was also observed in junctions with MAO barrier.[24] Most importantly, the lattice parameters of MAO can be tuned by changing the ratio of Mg and Al. The MAO barrier has already been used with other FM electrodes such as $Co_2FeAl$,[25,26] and $Co_2FeAl_{0.5}Si_{0.5}$.[27]

Several methods have been employed to fabricate MAO barriers in the past: plasma oxidation of a metallic Mg-Al layer,[27,16] electron beam evaporation,[28,29] RF magnetron sputtering,[30,31] and the post annealing of an Al/MgO bilayer.[32] Generally the PMA in the CoFeB/MgO junctions arises at the FeCo/MgO interface due to the proper hybridization of 2p orbitals of Oxygen with the 3d orbitals of Fe and Co,[33] and is also highly influenced by the choice of heavy metals on the other side of FM electrodes. So, in order to achieve the better PMA and the higher TMR ratio, the oxidation state of the barrier is critical. In plasma oxidation method, it is not easy to avoid the over-oxidation of FM electrodes. The RF magnetron sputtering is a viable method where a sizable TMR of 36% has been achieved in perpendicular MTJ (pMTJ) with CoFeB and MAO and oxygen deficiency was identified as a possible source of low TMR.[30] A two-step process with repetition of Mg-Al alloy deposition and post oxidation has also been studied.[34] At optimum oxidation state the cation-ordered spinel MAO structure was reported in contrast to the cation-disordered spinel structure from the one-step process.

In this study we utilized a three steps deposition process to fabricate MAO barriers by RF magnetron sputtering. This enables us to optimize the $O_2$ concentration in the MAO barrier, which results a TMR ratio above 60% at room temperature (RT), the highest value so far in pMTJs with MAO barriers to the best of our knowledge.

The film stack was deposited in ultra-high vacuum chamber with a base pressure of $10^{-9}$ Torr, under similar conditions of our previous studies.[35–38] The metallic layers including $Co_{20}Fe_{60}B_{20}$ (CoFeB) were deposited by the DC magnetron sputtering at 2 mTorr working pressure. The MAO barrier was deposited in three steps by the RF magnetron sputtering at 1.5 mTorr working pressure. The first MAO layer was deposited with a pure Ar for the time $t_1$, the second MAO layer was deposited by the reactive RF sputtering with $O_2$ and Ar for the time $t_2$ and the third MAO layer was deposited with a pure Ar for the time $t_3$. The fraction of reactive RF sputtering time, t is defined as the ratio of $t_2$ to the total RF sputtering time i.e., $t = t_2/(t_1+t_2+t_3)$. The Ar was fixed at 20 sccm for the whole RF sputtering process but the $O_2$ was varied from 0.14 sccm to 0.25 sccm which is equivalent to oxygen partial pressure of 10 µTorr to 19 µTorr. $t_1$, $t_2$, $t_3$ and the amount of $O_2$ were chosen to optimize the Oxygen concentration in the barrier but to minimize the over-oxidation of the adjacent CoFeB electrodes. The MTJs were annealed in a rapid thermal annealing setup under different conditions before the magnetic and transport measurement.[39,40]

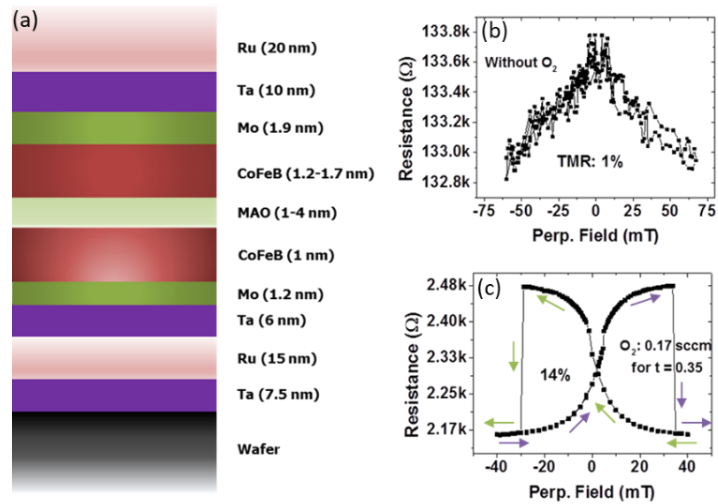

Figure 1: (a) Schematic of pMTJs. (b) TMR curve for the sample deposited without $O_2$. (c) TMR curve for the sample deposited with 0.17 sccm $O_2$ for t = 0.35. The arrows in (c) show the magnetic field sweeping direction [purple (green) towards the positive(negative) field direction].

Figure 1(a) shows the schematic of pMTJs with MAO barrier. Figure 1(b) shows the TMR curve for the sample with t = 0, i.e., the sample without oxygen and Figure 1(c) shows the TMR curve for the sample with t = 0.35. The TMR curve of the sample without $O_2$ only exhibits a weak magnetic field dependency, showing no sign of PMA in figure 1(b). However, even for a very small t, the PMA and the TMR ratio are significantly improved in figure 1(c). This is consistent with the understanding that PMA only exhibits in FM/Oxide layers with optimal interfacial oxidation states. The figures 1(b) and 1(c) represent the behavior of MAO-pMTJ for a wide range of CoFeB thicknesses from 1.2 nm to 1.7 nm before the optimization of $O_2$ concentration in the barrier. These results motivate us to further investigate the magnetic and transport properties of pMTJs with RF-sputtered MAO barrier.

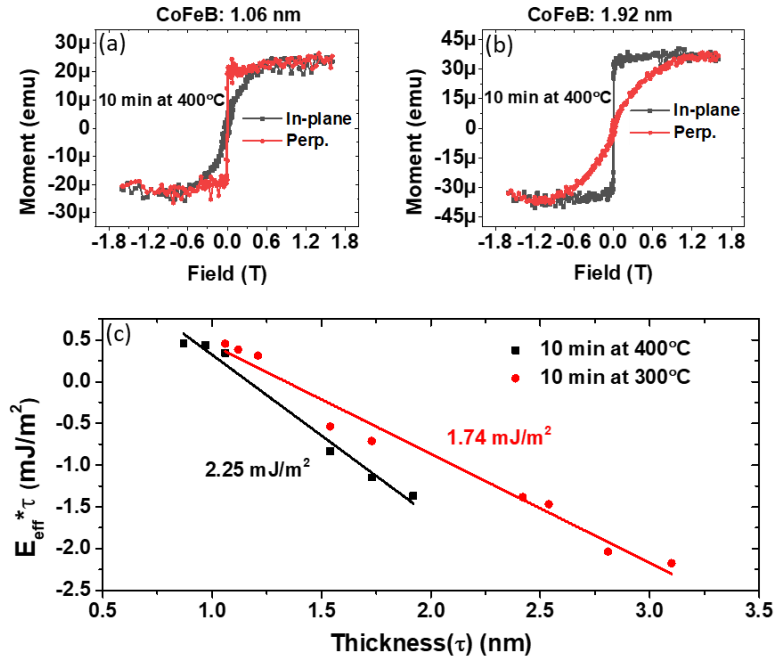

Figure 2: (a) and (b) Hysteresis loops of the samples with CoFeB thicknesses 1.06 nm and 1.92 nm respectively. These samples were annealed at 400˚C for 10 min. (c) The plot of $E_{eff}*\tau$ as a function of CoFeB thickness ($\tau$). Red curve is for the samples annealed at 300˚C for 10 min and Black curve is for the samples annealed at 400˚C for 10 min.

We first carried out the vibrating sample magnetometer (VSM) study on the magnetic properties of films. The sample structure used is Si/SiO2/Ta(7.5)/Ru(15)/Ta(6)/Mo(1.2)/CoFeB(0.75-3.25)/MAO(3)/Ta(5)/Ru(10), where the numbers in parentheses indicate the thickness in nanometers. The samples were annealed in an oxygen free glove box before performing the VSM measurement. During the annealing process B diffuses out from the CoFeB and is absorbed on the adjacent heavy metal layers and the barrier. Representative hysteresis loops are shown in figure 2(a) and 2(b) for the samples annealed at 400˚C for 10 min. Figure 2(a) shows the easy axis along the out-of-plane orientation for the sample with 1.06 nm CoFeB and figure 2(b) shows the easy axis along the in-plane orientation for the sample with 1.92 nm CoFeB. A series of VSM samples with CoFeB thicknesses ranging from 0.75 nm to 3.25 nm were fabricated and measured to extract the interfacial PMA energy density ($E_i$). The VSM samples were fabricated with $O_2$: 0.21 sccm for t = 0.67, this is the amount of $O_2$ where we have good PMA and higher TMR for the larger range of CFB thickness. For each sample, the saturation magnetization ($M_s$) and the anisotropy field ($H_k$) were determined from the hysteresis loops, which were used to find the effective PMA energy density ($E_{eff}$) as, $E_{eff} = \frac{1}{2}M_sH_k$. $E_{eff}$ can be expressed in terms of $E_i$ as, $E_{eff} = E_b - 2\pi M_s^2 + \frac{E_i}{\tau}$, where $E_b$ and $2\pi M_s^2$ are the bulk and shape anisotropy energy densities, respectively and $\tau$ is the thickness of CoFeB. $E_{eff}*\tau$ is plotted as a function of $\tau$ in figure 2(c), the y-intercept of which gives the $E_i$. In our study $E_i$ is found to be 1.74 mJ/m² for the samples annealed at 300°C for 10 min and 2.25 mJ/m² for the samples annealed at 400°C for 10 min. Previously $E_i$ of 1.30 mJ/m² was reported for the sputtered MAO-based samples annealed at 300°C.[30] However, their samples were not able to retain the PMA at 400°C. On the other hand, for the MgO-based similar structure annealed at 400°C for 10 min, PMA density of 1.74 mJ/m² was reported.[36] These results imply that the optimally oxidized MAO barrier supports a larger PMA compared to MgO barrier.

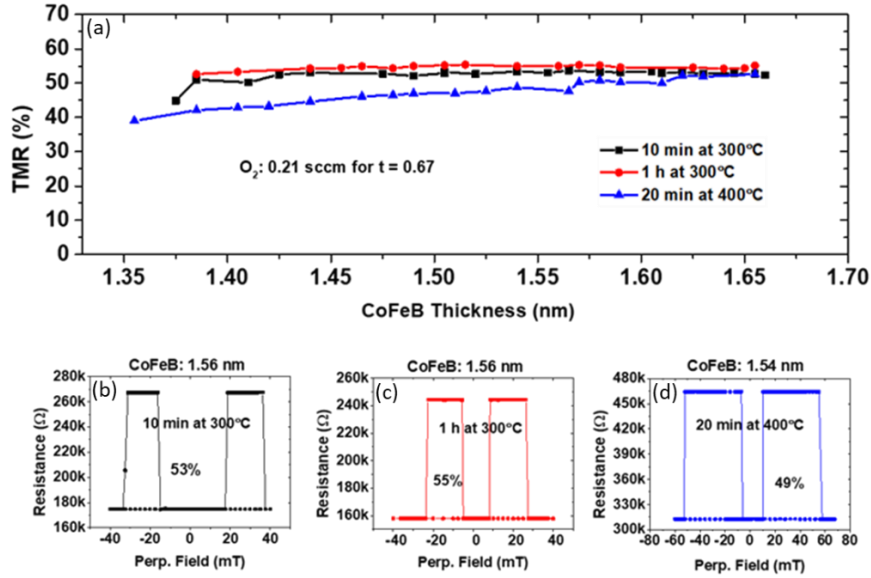

Figure 3: (a) The plot of TMR ratios vs. top CoFeB thickness (nm). Black curve is for the samples annealed at 300°C for 10 min, red curve is for the samples annealed at 300°C for 1 h and blue curve is for the samples annealed at 400°C for 20 min. (b), (c) and (d) represent the TMR curves for the samples annealed at 300°C for 10 min, 300°C for 1 h and 400°C for 20 min respectively.

Next, we discuss the transport properties in pMTJs. The sample has the structure of Si/SiO2/Ta(7.5)/Ru(15)/Ta(6)/Mo(1.2)/CoFeB(1)/MAO(1-4)/CoFeB(1.2-1.7)/Mo(1.9)/Ta(10)/Ru(20), where the numbers in parentheses indicate the thickness in nanometers. The MAO barrier was sputtered with 0.21 sccm $O_2$ with t = 0.67. The samples were then annealed at different conditions and measured at RT. The TMR ratios are plotted as a function of top CoFeB thicknesses in figure 3(a). The representative TMR curves for the pMTJs annealed at different conditions are shown in figures 3(b) – 3(d). It is seen that both the electrodes have the strong PMA with the sharp switching in all annealing conditions. The TMR ratio of about 53% was obtained for the pMTJs annealed at 300°C for 10 min, which increased slightly to 55% after annealing at 300°C for 1 h. The TMR ratio increased because the annealing improves the crystallization of the MTJ, which also plays an important role in determining the magnetic noise of the MTJs.[41] However, the TMR ratio dropped to about 50% after annealed at

400°C for 20 min. The insertion of thin Mo layer next to CoFeB prevents the Ta atoms diffusing to the barrier but allows the B atoms diffuse out through it.[36] Ta is known for the better B absorber,[7] so the right thickness of Mo layer should be chosen to take the advantage of both Ta and Mo. On the other hand if a thicker Mo layer is used, the highly texture Mo may promote unwanted texture for the CoFeB. In this study, we have used a relatively thicker Mo layer on both buffer (1.2 nm) and capping (1.9 nm) sides which might be the reason of lower TMR ratio after the 20 min annealing at 400°C. The PMA is still very strong with a flat anti-parallel plateau so, the reduced TMR after extended annealing at 400°C is more related to the change of transport, instead of magnetic properties of the MTJs. The relatively low TMR of these samples (compared to that of CoFeB/MgO junctions) indicates the MAO barrier still doesn't have the good crystalline structure as that of MgO.[37,42] The low TMR could be related to impurity induced tunneling.[43] At 300°C, the TMR ratios are almost similar for a wide range of top CoFeB thicknesses. However, at 400°C, the TMR ratios decrease continuously towards the thinner top CoFeB thicknesses. This could be the signature of over-oxidation of FM electrodes.

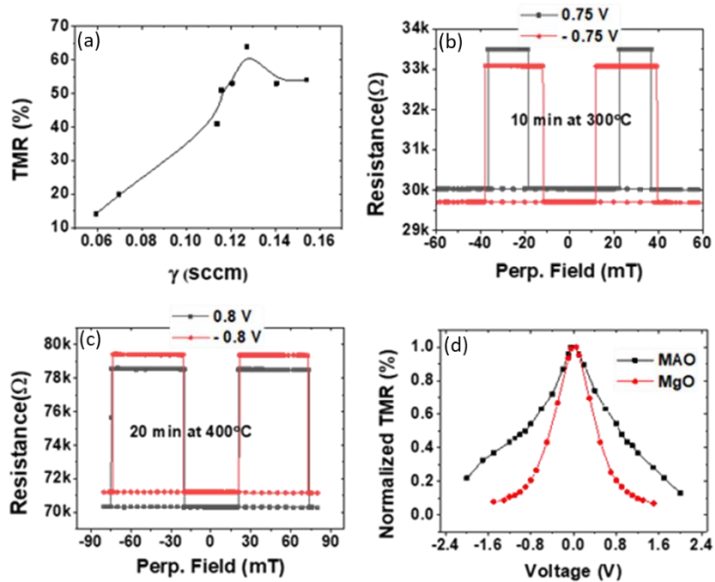

Figure 4: (a) Plot of TMR ratios as a function of γ, effective flow rate for the samples annealed at 300°C for 10 min. (b) and (c) represent the TMR curves for the samples annealed at 300°C for 10 min and 400°C for 20 min respectively. ± 0.75 V and ± 0.8 V biases were applied in (b) and (c) respectively. (d) Bias voltage dependence of normalized TMR ratio for the MTJ with MAO (black curve) and MgO (red curve) barrier for the samples annealed at 300°C for 10 min.

In order to better characterize the overall amount of oxygen used under different conditions, we define the effective flow rate, γ, as the product of *t* and the actual flow rate used during the reactive sputtering of MAO barrier. The TMR ratio as a function of γ is plotted in figure 4(a) for the samples annealed at 300°C for 10 min. The TMR ratio is continuously increased and reached above 60% at around γ = 0.13 sccm and dropped again and saturated at around γ = 0.15 sccm. The over oxidation could be the reason of dropping the TMR ratio for γ > 0.13 sccm. The slope on the left side of a peak indicates the oxygen deficiency in the sputtered MAO barrier. It is expected that by selecting the right γ, the barrier quality and hence the TMR ratio can be further improved. As mentioned earlier the lowest switching energy in MTJ is obtained by the VCMA effect.[10,11,44] In VCMA the external bias voltage creates an electric field in the barrier which is used to manipulate the interfacial magnetic anisotropy of the pMTJs. This effect can alter the coercive fields ($H_c$) of two ferromagnetic electrodes and gives a distinct TMR curves at different bias voltages.[45] The comparison of the VCMA effect after two different annealing conditions are shown in figures 4(b) and 4(c). The MAO samples used in figures 4(b), 4(c) and 4(d) were fabricated with $O_2$: 0.21 sccm for t = 0.67. In typical VCMA effect, negative bias voltage (corresponding to electrons flowing from top to bottom) decreases the $H_c$ of soft layer which is due to the re-distribution of electrons on different orbitals of the ferromagnetic electrodes.[46,47] A strong VCMA effect can be seen for the samples annealed at lower temperature in figure 4(b) however, there is no significant VCMA effect for the samples annealed at 400°C shown in figure 4(c). The over-oxidation of FM electrodes could be the reason of absence of VCMA effect for the sample annealed at 400°C, which explain why the TMR is lower after the 400°C annealing in Figure 3. Figure 4(d) demonstrates the bias voltage dependence of the normalized TMR ratio for the samples with MgO and MAO barriers. The sample with MgO barrier was identical to the sample with MAO barrier except the barrier. The $V_{half}$, bias voltage at which the TMR drops to half of the zero-bias value of MAO sample is +0.85 V (-1.20 V) for positive (negative) bias and that of MgO is 0.4 V (-0.42 V) respectively. These MgO $V_{half}$ values are close to the previously reported values.[7,48] For MAO sample the curve is asymmetrical with respect to the bias polarity which is due to the better crystallinity of top-interface. The $V_{half}$ values of sputtered MAO barriers are also comparable to the previously reported $V_{half}$ values of post oxidation-based MAO barriers.[15,16] These results indicate the energy efficient VCMA switching can be possibly obtained in pMTJ with MAO barriers.

In conclusion we successfully fabricated the pMTJs with sputtered MAO barrier using a three-step deposition process. Transport and magnetic properties of these MTJs were investigated after annealing under different conditions. TMR ratio above 60% was achieved by controlling the oxygen content in the barrier. A high $V_{half}$ of about 1.0 V was obtained, indicating promising voltage-related applications of MAO in pMTJs.

**Acknowledgements**

This contribution is dedicated to Chia-Ling Chien on his 80th birthday. Professor Chien is a pioneer in magnetism and superconductivity, who has inspired generations of researchers around the world. He has made numerous cutting-edge discoveries that are integral to today's understanding of spin-transport in solids. The senior author (W.W.) has the pleasure and privilege to have Chia-Ling as a mentor, friend, and colleague for more than 15 years, and is greatly inspired by his genius, dedication, and humor. This work was supported in part by Semiconductor Research Corporation through the Logic and Memory Devices program, by DARPA through the ERI program (FRANC), and by NSF through DMR-1905783 and ECCS-1554011.